\documentclass[aps,prl,showpacs, twocolumn, amsfonts, amsmath]{revtex4}

\usepackage{amsmath}
\usepackage{amssymb}
\usepackage{epsf}
\usepackage{epsfig}
\usepackage{graphicx}
\usepackage[english]{babel}
\usepackage[latin1]{inputenc}
\usepackage{rotating}
\usepackage{color}
\usepackage{slashed}
\usepackage{url}
\usepackage{textcomp}

\usepackage{txfonts}
\newcommand{\modified}[1]{{#1}}

\newcommand{\bea}{\begin{eqnarray}}
\newcommand{\eea}{\end{eqnarray}}

\renewcommand{\vec}[1]{{\bf #1}}

\newcommand{\eq}[1]{(\ref{eq:#1})}
\newcommand{\Eq}[1]{Eq.~(\ref{eq:#1})}

\newcommand{\Fig}[1]{Fig.~\ref{fig:#1}}

\begin{document}

\title{Charge separation in Reheating after Cosmological Inflation}

\author{Thomas Gasenzer}
\author{Boris Nowak}
\author{D\'enes Sexty}

\affiliation{Institut f\"ur Theoretische Physik,
             Ruprecht-Karls-Universit\"at Heidelberg,
             Philosophenweg~16,
             69120~Heidelberg, Germany}
\affiliation{ExtreMe Matter Institute EMMI,
             GSI Helmholtzzentrum f\"ur Schwerionenforschung GmbH, 
             Planckstra\ss e~1, 
             64291~Darmstadt, Germany}

\date{\today}

\begin{abstract}
  New aspects of parametrically resonant heating of a relativistic
  scalar $O(2)$-symmetric self-interacting field are presented.  This
  process is a candidate for reheating at the end of the
  early-universe epoch of inflation.  Although a model with a fully
  symmetric ground state is used, transient, metastable
  spontaneous symmetry breaking can be observed.  This manifests
  itself in the form of persistent regimes of opposite and, inside
  these, uniform charge overdensities separated by thin lines and
  walls similar to topological defects, in two and three spatial
  dimensions, respectively.  The configuration is found to correspond
  to \modified{a non-thermal} fixed point of the underlying
  dynamic equations for correlation functions which prevents thermalisation over an extended
  period of time. \modified{Our results establish a link between wave-turbulent 
  phenomena and the appearance of quasi-topological defects 
  in inflaton dynamics.}
\end{abstract}

\pacs{   98.80.Cq, 
11.10.Wx, 
47.27.E- 
 }

\maketitle

In cosmological models of the universe, reheating describes the epoch
starting at the end of inflation \cite{Allahverdi:2010xz}.  During
this epoch the potential energy of the inflaton field is redistributed
into a homogeneous and isotropic hot plasma of particle
excitations. These become a substantial part of the further expanding
universe.  Simple models describing reheating after inflation invoke
self-interacting scalar fields.  One of the popular scenarios involves
the parametrically resonant amplification of quantum fluctuations of
the macroscopically oscillating inflaton field.  The amplified modes
represent the emerging matter content of the universe
\cite{Kofman:1994rk,Traschen:1990sw}.  Various theoretical approaches
have been proposed to model reheating.  As both, the inflaton and the
amplified modes are strongly populated, classical field simulations
can be applied to describe their evolution
\cite{Khlebnikov:1996mc,Prokopec:1996rr,Tkachev:1998dc}.
Alternatively, Kadanoff-Baym dynamic equations, as derived from 2PI
effective actions in nonperturbative approximation, can describe the
resonant excitation and the ensuing thermalisation
\cite{Berges:2002cz}.  The exponentially fast excitation process is
followed by a slower equilibration, possibly with transient turbulent
behaviour transporting the energy deposited in the low-momentum modes of
the system to higher momenta.  Classical field simulations and
scaling solutions of kinetic equations were used to analyse possible
turbulent evolution during reheating and thermalisation
\cite{Micha:2002ey,Micha:2004bv}.  

\modified{Recent analytical predictions of new nonthermal fixed points of the field dynamics  \cite{Berges:2008wm} provided an extension of the above turbulence scenarios into the regime beyond perturbative quantum-Boltzmann approximations \cite{Berges:2008sr,Scheppach:2009wu}.
Being confirmed by numerical simulations \cite{Berges:2008wm,Berges:2010ez} these phenomena have triggered work in different areas \cite{Scheppach:2009wu,Carrington:2010sz,Nowak:2010tm}, in particular as they have the potential to strongly alter relaxation time scales.}
Here we demonstrate that these nonthermal fixed points are characterised by 
\modified{metastable configurations} of the underlying inflaton field \modified{with a soliton-like charge distribution}.
We study parametric resonance in a globally $O(2)$ or, equivalently, $U(1)$ symmetric relativistic scalar field theory. 
Shortly after the resonant excitations have set in we find spatial separation of charges.
Both, charge and anticharge overdensities become uniformly distributed within slowly varying regions which are separated by sharp boundary walls of grossly invariant thickness. 
These walls \modified{have a character similar to} topological defects and appear for generic initial conditions.
\modified{Metastable, non-topological defects have been studied, mostly in the context of non-linear single-scalar field theories, with or without a broken-symmetry ground state, where they are commonly termed oscillons \cite{Bogolyubsky:1976yu}.
In the present paper we connect quasi-stationary soliton-like charge separation in a two-component, $O(2)$-symmetric model to non-thermal stationary scaling solutions of quantum field dynamic equations which are known to exist for a far more general class of models.
In this way a link is established between wave turbulence phenomena as discussed, e.g., in \cite{Micha:2002ey,Micha:2004bv,Berges:2008wm,Berges:2008sr,Scheppach:2009wu,Berges:2010ez,Carrington:2010sz} and long-lived quasi-topological structures in the inflaton field. 
The very general concept of critical phenomena far from thermal equilibrium can provide a new way to classify nontrivial metastable field configurations. 
}

The action of the model considered here is given by
\bea
  S = \int d^d\!x\,dt \left\{{1\over 2} \left[ (\partial_{t} \varphi)^2 -  (\partial_i \varphi)^2 - m^2 \varphi^2\right] 
  - {\lambda \over 4! N} (\varphi^2)^2 \right\}
  \label{eq:ONmodel}
\eea
where $\varphi^{2}\equiv\varphi_{a}(x)\varphi_{a}(x)$, $(\partial_{i}\varphi)^{2}\equiv\partial_{i}\varphi_{a}(x)\partial_{i}\varphi_{a}(x)$, sums over $i=1,\dots,d$ and $a=1,\dots,N$, $N=2$, implied. 
To describe the dynamics of the scalar field $\varphi$ in the rapidly expanding universe one works in conformal coordinates in which the metric is defined through
%
\modified{$ 
  ds^2 = a(t)^2 ( dt^2 - dx^2 )
$,}
%
with a scale parameter $a(t)$ depending on the conformal time $t$.
We choose the bare mass scale to vanish, $m^{2}=0$, such that the equation of motion for the rescaled  field $\varphi ' = a \varphi$ reads (in $d=3$):
\bea 
  \left[\partial_t^2  - \Delta  
  + { \lambda\over 6 N }  {\varphi'}^{2} 
 - { \ddot a \over a }\right] \varphi'_{a} =0
\eea
%
%
\begin{figure*}[tb]
\begin{center}
\includegraphics[width=0.41 \textwidth]{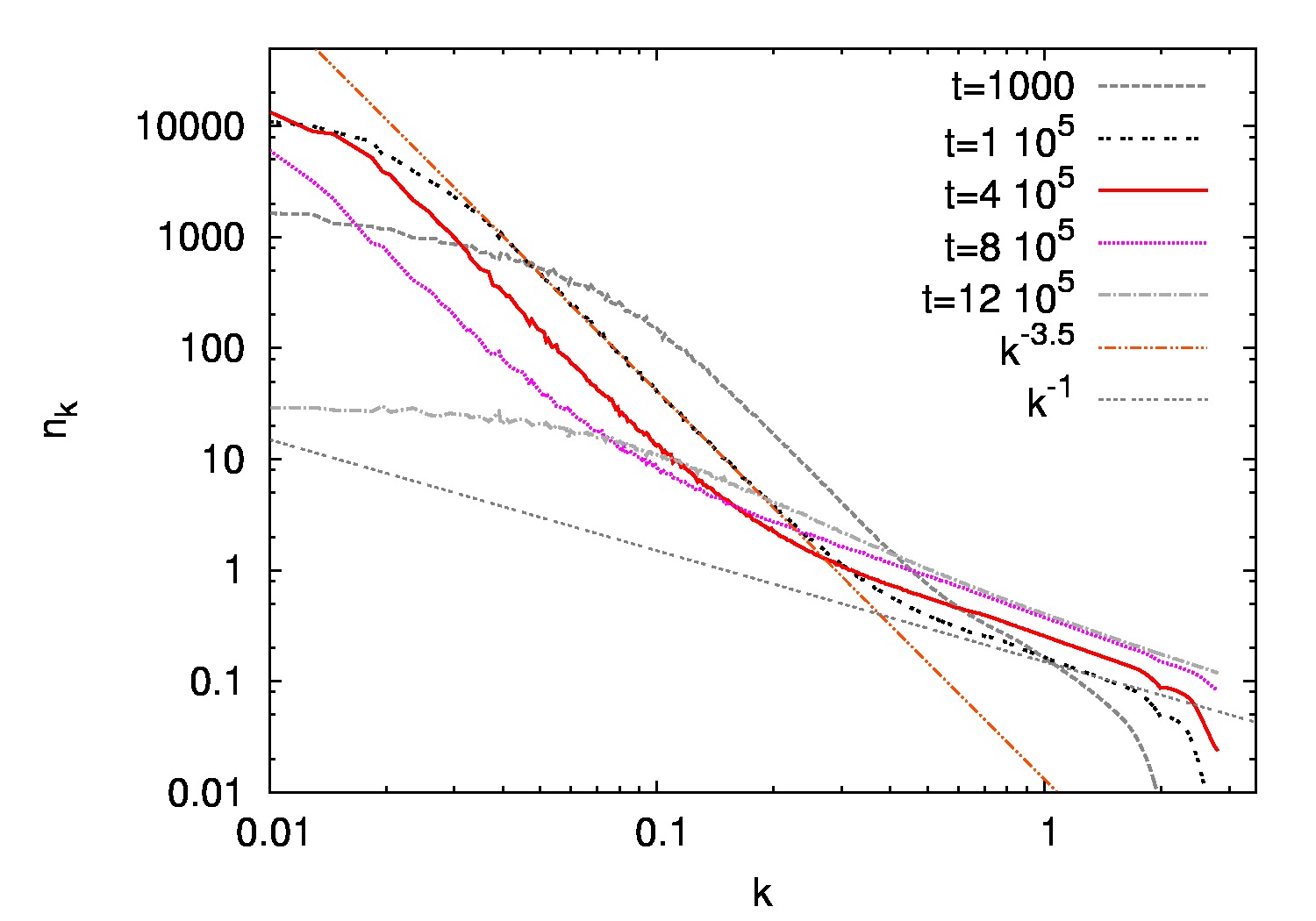}
\hspace*{0.07\textwidth}
\includegraphics[width=0.41 \textwidth]{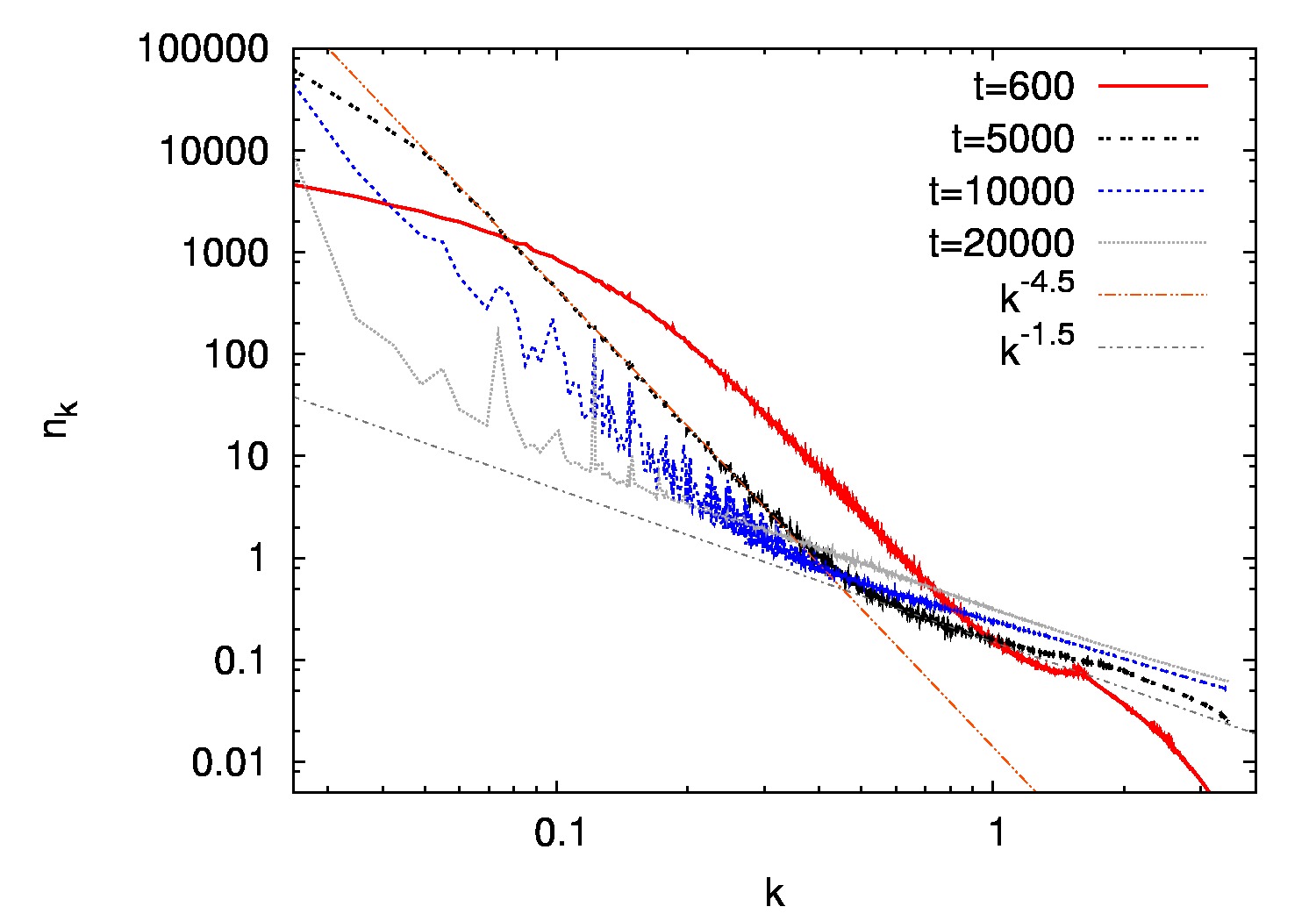}
\caption{(Color online) Occupation-number spectrum \eq{nkt} of the system in $d=2$ (left panel) and $d=3$ (right) spatial dimensions at different times $t$ (in lattice units). Averages were taken over $69$ (84)
runs on a $960^2$ ($256^3$) grid. 
\modified{The double-log scale exhibits the bimodal power-law $n(k,t)\sim k^{-\kappa}$. For $d=2$ we find $\kappa\simeq d+1.5$ in the infrared, and thermal scaling, $\kappa\simeq1$, for large $k$.}
For $d=3$ we also observe, at late times and intermediate momenta, the UV weak wave turbulence exponent $\kappa\simeq1.5$.
\modified{Note that the physical particle number spectrum is $6N n(k,t)/\lambda$.} 
See the main text for further details.
\label{fig:sp2d}}
\end{center}
\end{figure*}
%
During the parametric reheating epoch the universe is close to being radiation dominated \cite{Khlebnikov:1996mc}, such that the scale dependent term can be neglected, $\ddot a=0$.
Hence, the equation of motion for the rescaled field $\Phi = \varphi' /\varphi'_0 $, with $\varphi'_0 = \varphi'(t=0)$ the initial value of the inflaton field, can be approximated as
\bea \label{eq:eom}
 \left[\partial_t^2  - \Delta  + \Phi^2\right] \Phi_a = 0
\eea
where the coupling has been absorbed by the rescaling $\sqrt {\lambda/6N} \varphi'_0 x \rightarrow x$ and $\sqrt {\lambda/6N}\varphi'_0  t \rightarrow t$. 
Due to the global $O(2)$ symmetry of the action the charge-current density
\bea 
  j_{\mu}(x) = \Phi_1(x) \partial_{\mu} \Phi_2(x)  - \Phi_2(x) \partial_{\mu} \Phi_1(x). 
  \label{eq:charge}
\eea
is conserved, $\partial^{\mu}j_{\mu}=0$.

To induce parametrically resonant reheating, we start our simulation from a configuration where 
only the zero mode of the inflaton field is populated.
\modified{Fluctuating non-zero momentum modes act as seeds for the ensuing instabilities.} 
Choosing $m^{2}=0$ and $\lambda>0$, which corresponds to an equilibrium configuration in the symmetric phase, subsequent oscillation of the inflaton field induces parametrically resonant exponential growth of certain modes.
Scattering between these modes causes the entire spectrum to fill up.
Our simulations of \Eq{eom} were performed on a cubic space-time lattice \modified{of $L^{d}$ gridpoints} using leap-frog discretisation and periodical boundary conditions in $d=2$ and $3$ spatial dimensions.
The time evolution of the system after the instabilities have set in was found to be best described by a turbulent stage followed by  thermalisation on the largest time scales.
The turbulent phases feature universal scaling of the (solid-)angle-averaged momentum-space occupation numbers
\bea
  \label{eq:nkt}
  n(k,t) = {1 \over 4 L^d}\int d^{d-1}\Omega_{\vec k} \sqrt{\langle|\partial_{t}\Phi(\vec k,t)|^{2}\rangle\langle|\Phi(\vec k,t)|^{2}\rangle}.
\eea
\modified{From standard analytical studies based on quantum Boltzmann equations \cite{Zakharov1992a}, one expects to find, during these phases, the above radial occupation number to scale as $n(k,t)\sim k^{-\kappa}$ where the specific power-law exponent $\kappa$ reflects weak wave turbulence.
The distribution is approximately time independent, $n(k,t)\equiv n(k)$, for $k$ within the so-called inertial range, but in contrast to thermal equilibrium, energy is flowing through this range, between sources and sinks, establishing a non-equilibrium stationary state.
Different exponents $\kappa$ are possible, depending on whether the radial flow of energy is independent of $k$,  $P(k)\equiv P$, or the radial flow of (quasi-)particle number, $Q(k)\equiv Q$ \cite{Zakharov1992a}.
In summary, one finds the possible exponents  \cite{Zakharov1992a,Berges:2008wm,Scheppach:2009wu}}
\begin{equation}
  \label{eq:weakpower}
  \kappa_{P} = d + {3\over 2}(z-2) , \quad \kappa_{Q}=d+z-3.
\end{equation}
where $z$ is the homogeneity index of the dispersion $\omega(s\vec k)=s^{z}\omega(\vec k)$ of mode $\vec k$.
Note that \modified{in thermal equilibrium}  one expects $\kappa_{th}=z$, corresponding to the Rayleigh-Jeans law.
\modified{Given positive $\kappa$, occupation numbers grow large in the infrared (IR) and the Boltzmann perturbative approximation breaks down. 
In \cite{Berges:2008wm,Scheppach:2009wu}, an expansion of the two-particle irreducible (2PI) effective action to next-to-leading order in the inverse $1/N$ of the number of field components was used to extend the wave-turbulence analysis into the IR regime.
In this scheme, bubble-chain diagrams contributing to the self-energy in the dynamic Dyson equation are resummed, resulting in an effectively renormalised many-body coupling. 
With this, scaling exponents can also be derived in the IR regime:}
\begin{equation}
  \label{eq:strongpower}
  \kappa^\mathrm{IR}_{P} = d + 2 z  , \qquad \kappa^\mathrm{IR}_{Q} = d + z  
\end{equation}

\modified{As was first brought forward in \cite{Berges:2008wm} for the case of an $O(4)$ model, numerical simulations of parametric resonance confirm both, the IR and the UV scaling exponents. 
Corresponding results for the model studied here are shown in \Fig{sp2d} for $d=2$ and $d=3$ at different instances of time $t$. 
}

\modified{As was shown in \cite{Zakharov1992a}, for our model \eq{ONmodel}, in wave turbulence, either energy flows with rate $P$ towards larger $k$ or quasiparticles flow with rate $Q$ towards lower $k$.
In our simulations, the resonantly induced momentum distribution is observed to act, during the further evolution, as a source at intermediate momenta, $k\simeq0.1..0.3$, see Fig.~\ref{fig:sp2d}.}
Our model implies $z=1$ such that we expect from \eq{strongpower}, in the infrared, the exponent $\kappa^\mathrm{IR}_{Q}=d+1$, assuming a constant quasiparticle number flow towards the IR.
We note that in contrast to cases with smaller $1/N$ discussed in \cite{Berges:2008wm,Berges:2010ez} we find, in the IR, $\kappa\simeq3.5$ ($d=2$) and $4.5$ ($d=3$). 
\modified{In the UV, assuming a constant flow $P$ from intermediate to larger momenta one expects, from \eq{weakpower},} $\kappa^\mathrm{UV}_{P}=d-3/2$.
Numerically we find that in the UV the occupation number distribution approaches, at large times, thermal scaling $\kappa_{th}=1$.
For $d=3$ we also observe, at late times and intermediate momenta, the expected UV weak wave turbulence exponent $\kappa\simeq1.5$.

\begin{figure}[tb]
\begin{center}
\includegraphics[width=0.30 \textwidth]{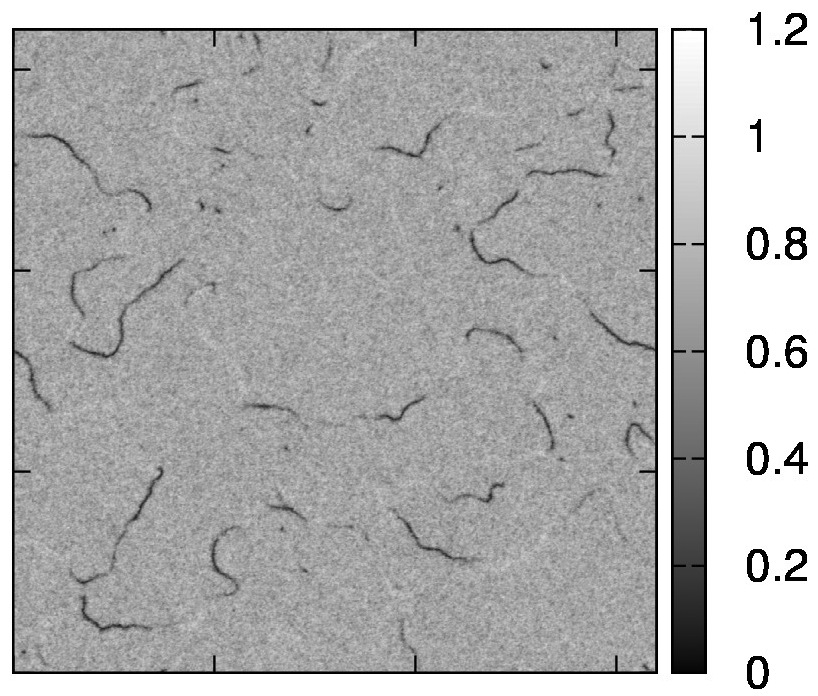}
\vspace*{-3ex}
\end{center}
\caption{Modulus $|\Phi(\vec x,t)|= [ \Phi_{1}^2(\vec x,t) + \Phi_2^2(\vec x,t)]^{1/2}$ at lattice time $t=9\times10^{3}$. 
Shown is its distribution over the two-dimensional $960^{2}$ lattice used in our simulations.
``Defect'' lines with substantially reduced $\rho=j_{0}$ (cf.~\eq{charge}),  contrast with a fairly constant bulk $\rho_\mathrm{bulk}\simeq0.7$.
\label{fig:kigyo2d}
}
\end{figure}

\begin{figure}[tb]
\begin{center}
\includegraphics[width=0.30 \textwidth]{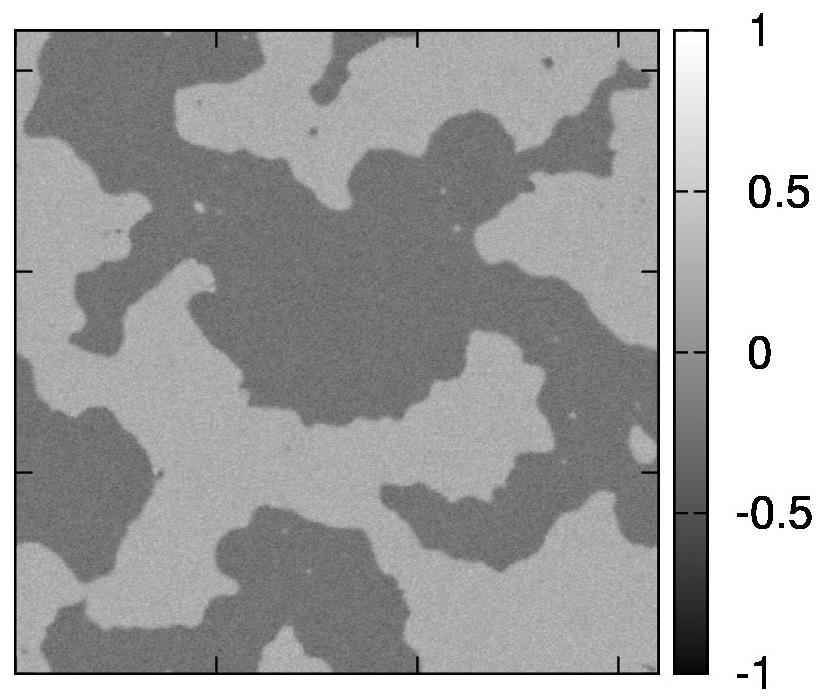}
\caption{Charge density $\rho(\vec x,t)=j_{0}(\vec x,t)$ (cf.~\eq{charge}), for the same configuration as shown in \Fig{kigyo2d}.
Note the evenly distributed charge overdensities with opposite sign, separated by lines which include the ``defect'' lines shown in \Fig{kigyo2d}. 
\label{fig:charge2d}}
\end{center}
\end{figure}

\begin{figure}[tb]
\begin{center}
\includegraphics[width=0.4 \textwidth]{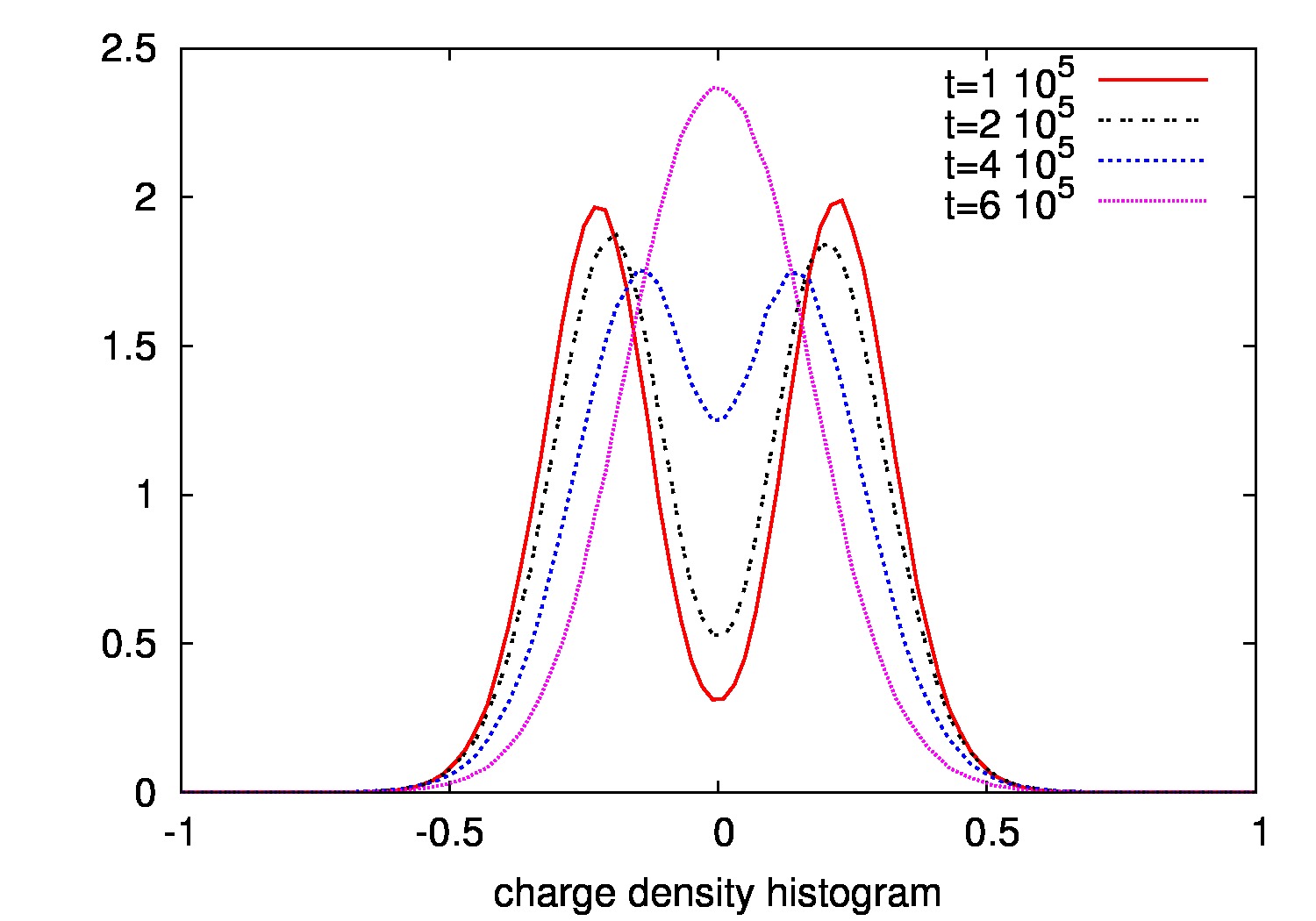}
\caption{Histogram of the charge density distribution of the system in $d=2$, at points of time for which \Fig{sp2d} shows the corresponding spectra.
We find that a clear charge separation goes together with the strong IR scaling with \modified{$\kappa=3.5$}.
\label{fig:hist2d}}
\end{center}
\end{figure}

\modified{Having recovered scaling behaviour} we present, in the following, our central
result: Looking at the real-space structure of the emerging
critical configuration we find patterns similar to topological defects
giving rise to quasi-stationary charge separation.  
\modified{We find a correspondence between the appearance of the strong scaling exponent and of the defect-separated charge patterns.}
In \Fig{kigyo2d}
we depict, for $d=2$, a typical real-space configuration in the
turbulent stage, plotting the modulus of the $O(2)$ scalar field,
$|\Phi(\vec x,t)|= [ \Phi_{1}^2(\vec x,t) + \Phi_2^2(\vec
x,t)]^{1/2}$.  Localised regions appear, specifically ``defect'' lines
where the absolute value of the field is much smaller than its
average.  \Fig{charge2d} shows the corresponding charge density 
$\rho(\vec x,t)=j_{0}(\vec x,t)$, Eq. \eq{charge}.  Clearly,
both uniform charge and anti-charge overdensities appear within
distinctly separated regions, showing only small fluctuations as
compared to their bulk values.  This separation of charges is
confirmed by the histogram of local charge densities 
shown in \Fig{hist2d}.  We found similar histograms for the case of
the $O(N)$ model \eq{ONmodel} with $N=3,4$ which we will
study in more elsewhere.
%
%
%

Moreover, we see that the ``defect'' lines in \Fig{kigyo2d} are lying along the boundary limits between the regions of opposite charge.
Note that only part of the boundaries is clearly visible as ``defect'' lines.
Following the time evolution of the defect lines \cite{ONVideos} we find that they appear and disappear with a period of $T/2\simeq6$ determined by the effective mass $m_\mathrm{eff}=2\pi/T$ of the mean field theory, at every point along the boundaries between the oppositely charged domains.
%
%
The explanation for the appearance of these lines is as follows:
The charge density $\rho$ can be interpreted as field angular momentum in the $N=2$ dimensional field space. 
Hence, $d$-dimensional configuration space can be divided into regions according to the sign of $\rho$.
On the lines separating these regions the local field variable has neighbours of opposite charge which therefore possess opposite ``circular polarisation'' in field space. 
Thus, at a particular point on the boundary the rotations add to a linear oscillation of the field through zero, giving rise to spatially and temporally oscillating defect lines.
While such an evolution could similarly occur in a free theory ($\lambda=0$) interactions are responsible for rendering the transition between the opposite charges sharp as can be seen in \Fig{kigyo2d}.

\begin{figure}[tb]
\begin{center}
\ \\[-2ex]
\begin{minipage}[b]{0.29\textwidth}
\includegraphics[width=0.49 \textwidth]{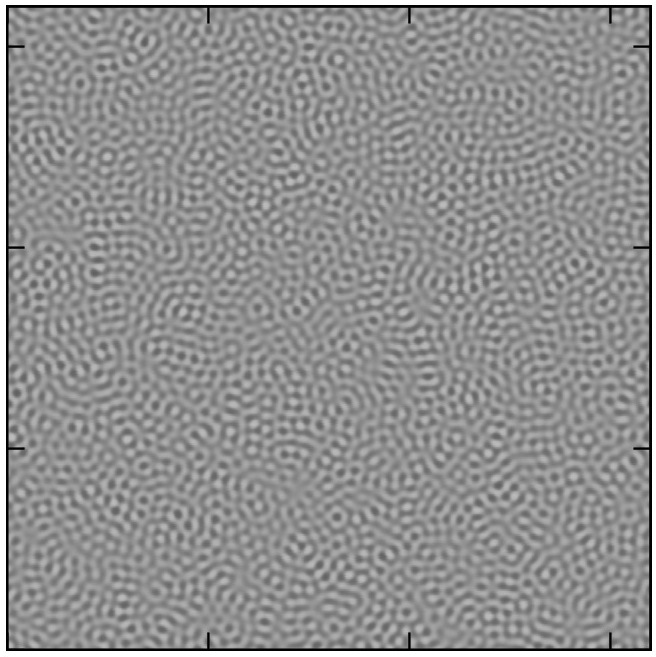}
\includegraphics[width=0.49 \textwidth]{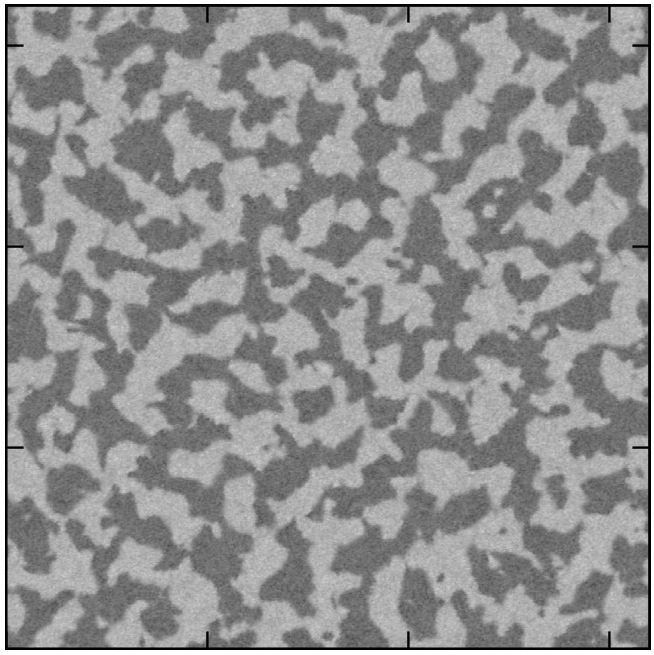}\ \\
\includegraphics[width=0.49 \textwidth]{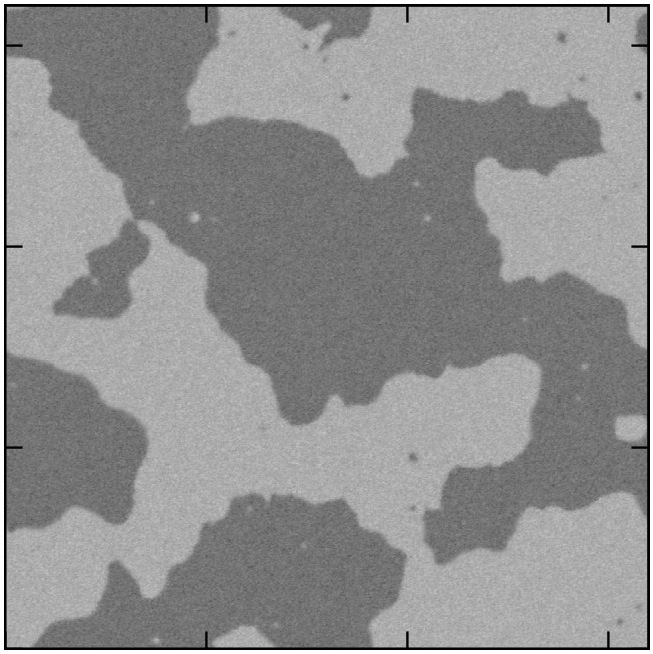}
\includegraphics[width=0.49 \textwidth]{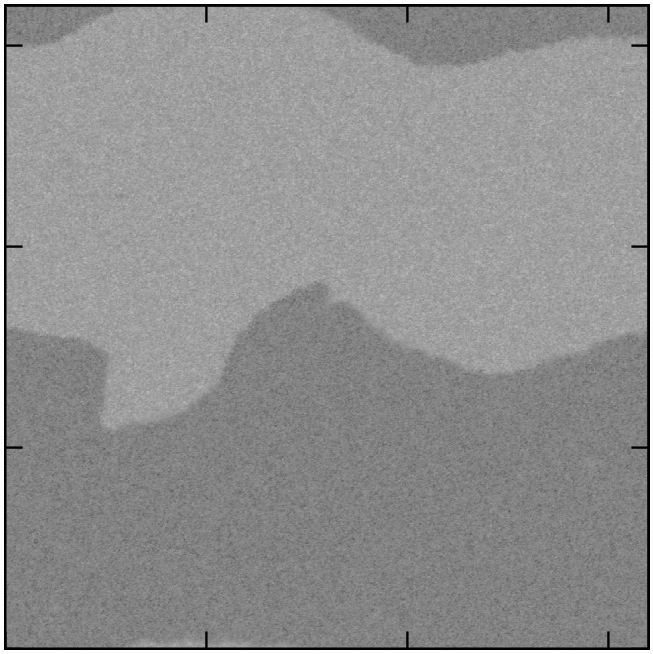}
\end{minipage}
\includegraphics[width=0.062 \textwidth]{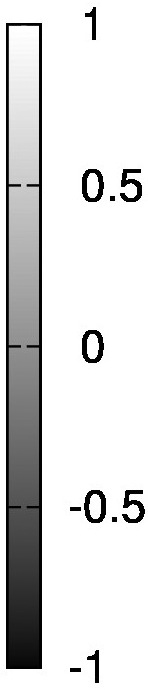}
\caption{Charge density $\rho(\vec x,t)=j_{0}(\vec x,t)$ (cf.~\eq{charge}) for the same run as shown in \Fig{charge2d} at the times $t=300$ (upper left), $t=10^{3}$ (upper right), $t=10^{4}$ (lower left), $t=6\times10^{4}$ (lower right).
The scale of the largest structures is found not to depend on the total lattice size. 
\label{fig:timelinecharge2d}}
\end{center}
\end{figure}
%
Following the evolution of the system starting from the homogeneous initial configuration we find the defect lines to appear once the instability has saturated.
The lines can reconnect with each other while they propagate slowly through configuration space, such that the number of spatially connected charged regions changes in time.
On the average, the total length of the defect lines as well as the number of connected regions is found to decrease with time, see \Fig{timelinecharge2d}.
During the late stage of the evolution the separation of opposite charges gradually disappears while the system finally thermalises.
This can also be seen in the histogram in \Fig{hist2d} where the central well gradually disappears while the spectrum $n(k,t)$ approaches thermal equilibrium, see \Fig{sp2d}.
We note that the sharp step between the opposite charges remains (\Fig{timelinecharge2d}, bottom right panel).
This, as well as the histogram in \Fig{hist2d} suggest that topological defects occur according to the homotopy group of the charge density space $C$ instead of the order parameter space.
Although $C$, for the $O(2)$ model, equals the real numbers, our results show, that  the system remains, for a long time, in a subspace which is topologically equivalent to $S_0= \{-1,1\}\in C$.
Hence, one expects defect lines (walls) to separate regions corresponding to the two unconnected parts of $C$.

\begin{figure}[tb]
\begin{center}
\ \\[-3ex]
\includegraphics[width=0.4 \textwidth]{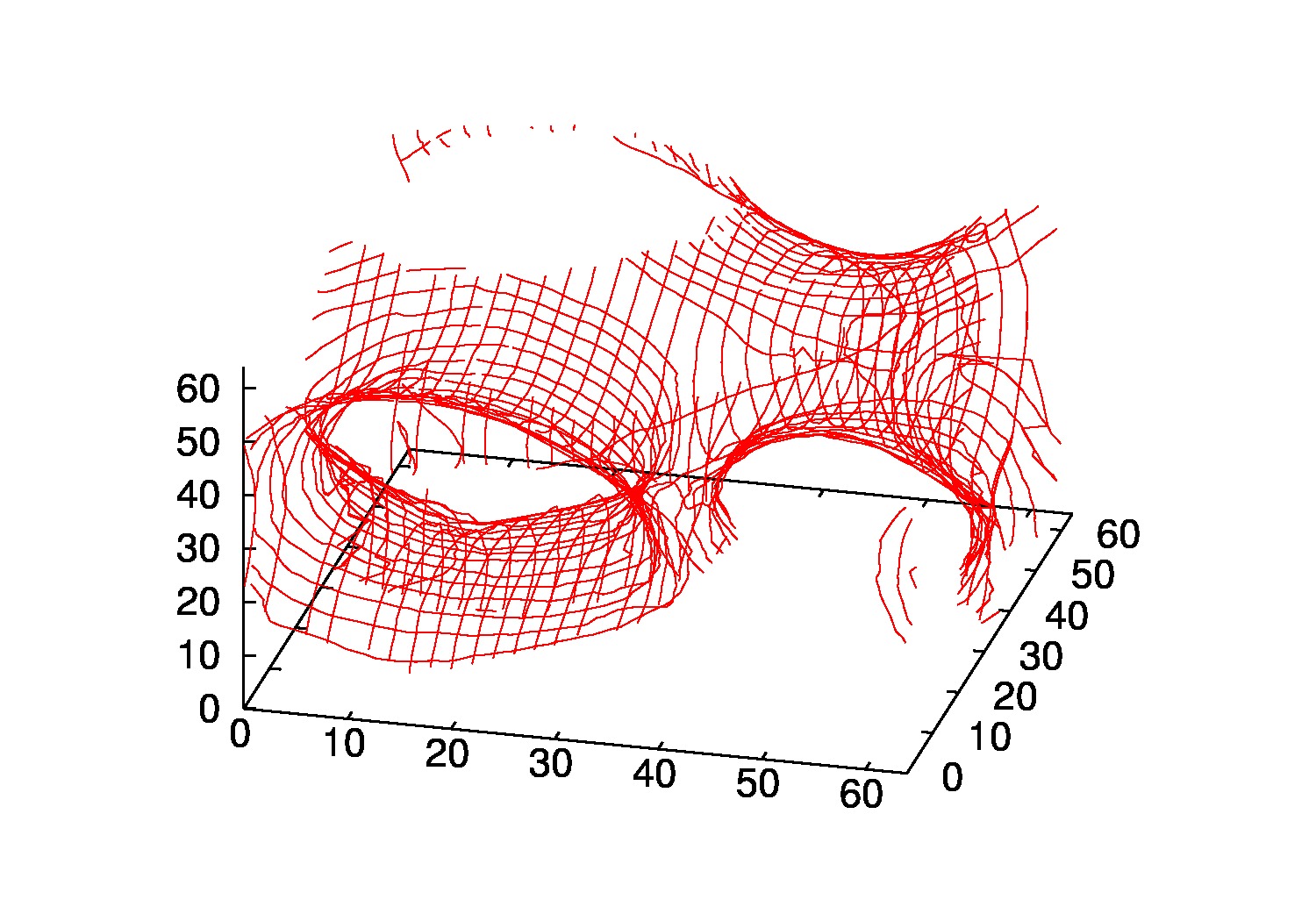}\ \\[-3ex]
\caption{Bubbles of approximately constant and opposite charge separated by  thin walls on a $64^3$ lattice at $t=1500$. Plotted are surface lines which 
are fitted to lattice points having $ |\Phi|^2 < 0.05 $.
\label{fig:bubbles}}
\end{center}
\end{figure}

For $d=3$, starting from an equivalent initial configuration, we find bubbles of opposite
charge separated by thin walls (\Fig{bubbles})  \cite{ONVideos} as well as charge density histograms similar to the case $d=2$.
These bubbles are seen to appear in coincidence with the strong IR scaling  in the spectra \modified{(\Fig{sp2d}, right panel)}.

We emphasise that the structures found are distinctly different from the known topological defects appearing in $O(2)$ theories such as vortices in $d=2$ and vortex strings in $d=3$ which one obtains in the case of equilibrium symmetry breaking, i.e., $m^{2}<0$ \cite{Tkachev:1998dc}. 
The bubble walls reported here form for vanishing or positive mass squared. 
%
%

Considering other potential cosmological consequences of our findings, notice that, in a different context, the $U(1)$-symmetric scalar field theory can also be understood as an ingredient of Affleck-Dine baryogenesis \cite{Affleck:1984fy}, where the scalar field carries a baryon charge and is thought to be a supersymmetric partner of standard-model fields. 
We emphasize that, in contrast to such models, there is no symmetry breaking implemented in our model from the outset.
The breaking rather occurs as a dynamical, transient but quasistationary effect.

To summarize, we have studied parametric reheating in an $O(2)$ symmetric scalar field theory in two and three spatial dimensions.
\modified{We have found structures similar to \modified{(quasi-)} topological defects to emerge on intermediate time scales, breaking charge symmetry locally while the classical action implies an equilibrium configuration without symmetry breaking. 
These structures were found to appear in direct correspondence with strong wave-turbulent scaling behaviour of occupation numbers in the IR regime of low momenta.
This establishes a strong link between wave-turbulence and quasi-stationary defect-like pattern formation of inflaton dynamics in scenarios of reheating. 
It is an interesting question whether similar structures exist in more complicated theories, such as gauge theories.}
\acknowledgements \noindent
The authors would like to thank J. Berges, J. M. Pawlowski, and M. G. Schmidt for inspiring discussions.  
They acknowledge support by the DFG (GA 677/7, 8), the University of Heidelberg (FRONTIER), the Alliance Program of the Helmholtz Association (HA216/EMMI), by BMBF and MWFK Baden-W\"urttemberg (bwGRiD cluster), and in part by the National Science Foundation under Grant No. PHY05-51164. The authors thank KITP Santa Barbara for its hospitality.


\begin{thebibliography}{18}
\expandafter\ifx\csname natexlab\endcsname\relax\def\natexlab#1{#1}\fi
\expandafter\ifx\csname bibnamefont\endcsname\relax
  \def\bibnamefont#1{#1}\fi
\expandafter\ifx\csname bibfnamefont\endcsname\relax
  \def\bibfnamefont#1{#1}\fi
\expandafter\ifx\csname citenamefont\endcsname\relax
  \def\citenamefont#1{#1}\fi
\expandafter\ifx\csname url\endcsname\relax
  \def\url#1{\texttt{#1}}\fi
\expandafter\ifx\csname urlprefix\endcsname\relax\def\urlprefix{URL }\fi
\providecommand{\bibinfo}[2]{#2}
\providecommand{\eprint}[2][]{\url{#2}}

\bibitem{Allahverdi:2010xz}
  R.~Allahverdi, R.~Brandenberger, F.~Y.~Cyr-Racine and A.~Mazumdar,
  Ann.\ Rev.\ Nucl.\ Part.\ Sci.\  {\bf 60}, 27 (2010).

\bibitem{Kofman:1994rk}
  L.~Kofman, A.~D.~Linde and A.~A.~Starobinsky,
  Phys.\ Rev.\ Lett.\  {\bf 73}, 3195 (1994).

\bibitem{Traschen:1990sw}
  J.~H.~Traschen and R.~H.~Brandenberger,
  Phys.\ Rev.\  D {\bf 42}, 2491 (1990).

\bibitem{Khlebnikov:1996mc}
  S.~Y.~Khlebnikov and I.~I.~Tkachev,
  Phys.\ Rev.\ Lett.\  {\bf 77}, 219 (1996).

\bibitem{Prokopec:1996rr}
  T.~Prokopec and T.~G.~Roos,
  Phys.\ Rev.\  D {\bf 55}, 3768 (1997).

\bibitem{Tkachev:1998dc}
  I.~Tkachev, S.~Khlebnikov, L.~Kofman and A.~D.~Linde,
  Phys.\ Lett.\  B {\bf 440}, 262 (1998).

\bibitem{Berges:2002cz}
  J.~Berges and J.~Serreau,
  Phys.\ Rev.\ Lett.\  {\bf 91}, 111601 (2003).

\bibitem{Micha:2002ey}
  R.~Micha and I.~I.~Tkachev,
  Phys.\ Rev.\ Lett.\  {\bf 90}, 121301 (2003).

\bibitem{Micha:2004bv}
  R.~Micha and I.~I.~Tkachev,
  Phys.\ Rev.\  D {\bf 70}, 043538 (2004).

\bibitem{Berges:2008wm}
  J.~Berges, A.~Rothkopf and J.~Schmidt,
  Phys.\ Rev.\ Lett.\  {\bf 101}, 041603 (2008).

\bibitem{Berges:2008sr}
  J.~Berges and G.~Hoffmeister,
  Nucl.\ Phys.\  B {\bf 813}, 383 (2009).

\bibitem{Scheppach:2009wu}
  C.~Scheppach, J.~Berges and T.~Gasenzer,
  Phys.\ Rev.\  A {\bf 81}, 033611 (2010).

\bibitem{Berges:2010ez}
  J.~Berges and D.~Sexty,
  Phys.\ Rev.\  D {\bf 83}, 085004 (2011).

\bibitem{Carrington:2010sz}
  M.~E.~Carrington and A.~Rebhan,
  arXiv:1011.0393 [hep-ph].

\bibitem{Nowak:2010tm}
  B.~Nowak, D.~Sexty and T.~Gasenzer,
  Phys.\ Rev.\  B {\bf 84}, 020506(R) (2011).
  
\bibitem{Bogolyubsky:1976yu}
  I.~L.~Bogolyubsky, V.~G.~Makhankov,
  Pisma Zh.\ Eksp.\ Teor.\ Fiz.\  {\bf 24}, 15 (1976);
%
  M.~Gleiser,
  Phys.\ Rev.\  D {\bf 49}, 2978 (1994);
and, e.g.,  
  M.~A.~Amin, R.~Easther, H.~Finkel,
  JCAP {\bf 1012}, 001 (2010),
and refs.~therein.
  

\bibitem[{\citenamefont{Zakharov et~al.}(1992)\citenamefont{Zakharov, {L'vov},
  and Falkovich}}]{Zakharov1992a}
\bibinfo{author}{\bibfnamefont{V.~E.} \bibnamefont{Zakharov}},
  \bibinfo{author}{\bibfnamefont{V.~S.} \bibnamefont{{L'vov}}},
  \bibnamefont{and}
  \bibinfo{author}{\bibfnamefont{G.}~\bibnamefont{Falkovich}},
  \emph{\bibinfo{title}{Kolmogorov Spectra of Turbulence I: Wave Turbulence}}
  (\bibinfo{publisher}{Springer-Verlag, Berlin}, \bibinfo{year}{1992}).

\bibitem[{ONV()}]{ONVideos}
\bibinfo{journal}{For videos of the evolution see\newline
  {\small\texttt{http://www.thphys.uni-heidelberg.de/{\texttildelow}smp/sexty/videos}}}

\bibitem{Affleck:1984fy}
  I.~Affleck and M.~Dine,
  Nucl.\ Phys.\  B {\bf 249}, 361 (1985).

\end{thebibliography}

\end{document}